\documentclass[12pt]{article}
\title{Doubly Special Relativity versus $\kappa$-deformation
of relativistic kinematics}
\author{by\\
Jerzy Lukierski$^*$ \\ Institute for Theoretical Physics, University of
Wroc\l{}aw,\\ pl. Maxa Borna 9, 50-204 Wroc\l{}aw, Poland\\ and \\ Anatol
Nowicki\thanks{supported by KBN grant 5P03B05620.}\\ Institute of Physics,
University of Zielona G\'ora,\\ pl. S\l{}owia\'{n}ski 6, 65-069 Zielona G\'ora,
Poland }

\def\r#1{(\ref{#1})}

\def\bel{\begin{equation}\label}
\def\ee{\end{equation}}
\def\bea{\begin{array}}
\def\ea{\end{array}}
\newcounter{lit}
\newenvironment{lit}
               { \stepcounter{equation}
             \setcounter{lit}{\value{equation}}
             \setcounter{equation}{0}
             
                }
           { \setcounter{equation}{\value{lit}}
             
                }

\def\0{^{(0)}}
\def\cop{\Delta}
\def\tens{\otimes}
\begin{document}
%\begin{titlepage}
\date{}
\maketitle
\begin{abstract}
We argue that recently proposed by Amelino-Camelia et all [1,2] so-called
doubly special relativity (DSR), with deformed boost transformations identical
with the formulae for $\kappa$-deformed kinematics in bicrossproduct basis is a
classical special relativity in nonlinear disguise. The choice of symmetric
composition law for deformed fourmomenta as advocated in [1, 2] implies that
DSR is obtained by considering nonlinear fourmomenta basis of classical
Poincar\'{e} algebra and it does not lead to noncommutative space-time. We also
show how to construct large two classes of doubly special relativity theories -
generalizing the choice in [1,2] and the one presented by Magueijo and Smolin
[3]. The older version of deformed relativistic kinematics, differing
essentially from classical theory in the coalgebra sector and leading to
noncommutative $\kappa$-deformed Minkowski space is provided by quantum
$\kappa$-deformation of Poincar\'e symmetries.
\end{abstract}

%\end{titlepage}

\section{Introduction}
The $\kappa$-deformation of Poincar\'e symmetries in the form of dual pair of
Hopf algebras describing respectively quantum deformations of $D=4$ Poincar\'e
algebra and $D=4$ Poincar\'e group [4-11] was proposed in order to introduce
modification of relativistic kinematics at extremely high energies. Recently,
in the search for phenomenological high energy effects of quantum gravity (see
e.g. [12-15]) there were also put forward the generalizations of special
relativity theories with two observer-independent parameters -- light velocity
$c$ and Planck length $L_p$ -- under the name of doubly special relativity
(DSR) [1,2, 16-17]. In concrete realization of DSR framework the basic formulae
coincide with the ones following from the algebraic sector of
$\kappa$-deformation of Poincar\'e algebra in bicrossproduct basis [8-10], with
the assumption that the mass-like deformation parameter $\kappa$ is equal to
the Planck mass $m_p$. In particular the basic characteristic of DSR -- the
deformed dispersion relation, described in [1] as ,,key characteristic of
DSR1\footnote{In [1] there are considered two DSR theories: DSR1 and DSR2 --
the second one based on [3]. We consider different versions of DSR in Sect. 2},
both conceptually and phenomenologically'' \bel{1} m^2=\left(\frac{\sinh
\frac12L_pE}{\frac12L_p}\right)^2- \vec p{\,}^2 e^{L_pE} \ee is exactly the
formula for $\kappa$-deformed mass-shell condition in bicrossproduct basis
(assuming $c=1$); also the differential realizations of the generators of boost
transformations in DSR, calculated in [17], can be found in [9].

The aim of this note is however not to intervene in the procedure of
referencing to the previous achievements. In fact we are happy that others are
applying and developing our ideas, even if this is done without providing
proper links with already established results. Here we would like to clarify
the notions of DSR from the point of view of Hopf algebras, its coalgebraic
structure and its relation with $\kappa$-deformation of Poincar\'e symmetries
[5-12].

A flag property of DSR is the appearance of limiting momentum achieved at
infinite  energy described by Planck mass, which can not be overpassed if we
change reference frame. The main issue is which properties will have an object
composed of two constituents having the above described limiting momentum. We
can have in principle two possible solutions:
\begin{enumerate}
\item[i)] Only single objects (one can call them ,,Plancktons'') do have
the property of limiting momentum. In such a case the Planck mass is just a
constant parameter in the theory entering deformed boost transformations,
related possibly with some quantum gravity-related quanta.

\item[ii)] The property of existence of maximal limiting momentum is valid in
 the composition process of the momenta.
Similarly like in classical special relativity theory one has the relativistic
addition law of velocities \bel{2} v_{12}=v_1 \dot + v_2 \Rightarrow c\dot + c
=c \ee one should have an addition law $p_{12}= p_1\dot +p_2$ providing the
existence of the same limiting momenta for $p_{12}$. It appears that in such a
case the deformed symmetries should be described as a Hopf algebra with
homomorphic coproducts, what ensures that the boost transformations for the
constituents $p_i$ ($i=1,2$) and of total momentum $p_{12}$ shall have the same
form, in particular the same limiting momentum.
\end{enumerate}
In [1,2] the authors propose (in linear approximation) the following
composition law for energy-momenta in DSR1 theory:
\footnote{In comparision with [1] we inserted in formula  \r{4}
the factor $\frac1c$ consistent with the relation $L_p=\frac1{M_pc}$.}
\bel{3} E_1 + E_2 - c L_p
 \vec {p}_1  \vec{p}_2 \ = \  E_1{\,'} + E_2{\,'} - c L_p \vec{p}_1{\,'}
\vec{p}_2{\,'}
\ee
\bel{4}
\vec{p}_1 +\vec{p}_2 - \frac {L_p}{c}  (E_1
\vec{p}_2 +E_2 \vec{p}_1) \ = \ \vec{p}_1{\,'} +\vec{p}_2{\,'}
- \frac{ L_p}c (E_1{\,'} \vec{p}_2{\,'} + E_2{\,'} \vec{p}_1{\,'})
\ee
These composition laws disclose the Hopf algebra  structure of the
theory proposed in [1,2]: the relations $(3,4)$ describe the part
linear in $L_p$ of the coproduct obtained by nonlinear
transformation of the fourmomenta in classical relativistic
theory, i.e. the basic structure remains Einsteinian. Such
theories with nonlinear symmetric coproduct for energy were
considered [18] just after appearance of $\kappa$-deformations of
relativistic symmetries in order to use the nonlinear composition
law as a tool to describe the dark matter effect, but were rather
abandoned in favour of  deformed theories providing abelian
addition law for energy.

In this note firstly in Sect. 2 we shall show that the deformed
boost transformations from [1,2,17] are indeed obtained by
nonlinear transformation of momentum generators in classical
relativistic theory; besides we shall derive the composition laws
(3,4) for the energy and threemomentum by performing the same
nonlinear transformation of primitive coproduct for classical
fourmomenta (we recall that primitive coproduct describe classical
abelian fourmomentum composition law). By generalizing the
deformed mass-shell condition (1) we shall describe two large
classes of nonlinearly transformed classical theories,
generalizing DSR1 type (see [1,2]) and DSR2 type (see [3]) of
theories.

In Sect. 3 we shall consider the composition law described by the
coproduct of $\kappa$-deformed Poincar\'e algebra in
bicrossproduct basis. In particular following result present in
[9] we shall remind why the method of calculating boost formulas,
used in [17], is also valid in $\kappa$-deformed quantum-group
framework. Further as new result we shall show the consistency of
finite boost formulae with $\kappa$-deformed nonabelian addition
law for the momenta.

In Sect. 4, we present general discussion. We argue that the
reason for selecting the deformed mass-shell condition $(1)$ lies
in quantum structure of coproduct -- in the presence of symmetric
coproduct such a choice is not distinguished at all. We shall
show that under quite plausible assumptions there are only two
choices of momentum coproduct (addition law of momenta): either
one chooses a classical one (symmetric) or the one provided by
$\kappa$-deformation. Unfortunately, quantum group structure is
quite rigid and can not be cheated.

\section{Classical relativistic symmetries in nonlinear dis\-guise and DSR
theories}

Let us consider standard relativistic symmetries, described by the following
classical Poincar\'e Hopf algebra\footnote{In  this paper we consider the
hermitean generators of symmetry algebras.} ($g_{\mu\nu} = (-1,1,1,1)$):
\begin{lit}
\begin{eqnarray}
\label{5a}[P\0_\mu,P\0_\nu]&=&0\\
\label{5b}{}[M\0_{\mu\nu},P\0_\rho]&=&i(g_{\mu\rho} P\0_\nu -
g_{\nu\rho}P\0_\mu)\\
\label{5c}{}[M\0_{\mu\nu},M\0_{\rho\tau}]&=&i(g_{\mu\rho}M\0_{\nu\tau} +
g_{\nu\tau}M\0_{\mu\rho} - g_{\mu\tau}M\0_{\nu\rho} - g_{\nu\rho}M\0_{\mu\tau})
\end{eqnarray}
\end{lit}
with primitive coproduct \bel{6} \cop\0(I_A) \ = \ I_A\tens 1 + 1 \tens I_A
\qquad I_A = P\0_\mu ,M\0_{\mu\nu} \ee and bilinear mass Casimir representing
relativistic dispersion formula\\ ($E\0 = cP\0_0$) \bel{7}
C_2(E\0,\vec{P}^{(0)}) \ = \ (E\0)^2 - c^2 (\vec P\0)^2 \ = \ inv \ \equiv \
m_0^2 c^4 \ee Let us introduce by means of the nonlinear relations new
fourmomenta $P_\mu = (E, \vec{P})$ (see also [19]) \bel{8} E\0 \ = \ \kappa c^2
f\left(E,\vec {P}{}^2\right)\qquad
 \vec P\0 \ = \ \vec{P} g\left(E,\vec{P}{}^2\right)\ee
where the masslike parameter $\kappa$ describes the measure of nonlinearity
\footnote{If we assume for the functions $f(x,y)$ and $g(x,y)$ the Taylor
expansion at $x=0$, the linear term, proportional to $1/\kappa$ describes the
leading nonlinear correction for large $\kappa$.}. One assumes that \bel{9}
f(0,0) \ = \ 0 \qquad \frac{\partial f}{\partial x}(0,0) \ = \  g(0,0) \ = \ 1
\ee what provides in the limit $\kappa \to \infty$ the relations $E = E\0$,
$\vec{P} = \vec P\0$. In new variables $E$, $\vec{P}$ the relation (7) looks as
follows \bel{10} \kappa^2 c^2 f^2\left(E,\vec{P}{}^2\right) - \vec{P}{}^2
g^2\left(E,\vec{P}{}^2\right) \ = \ inv \ = \ m_0^2c^2 \ee One can introduce
two types of nonlinear transformations \r{8}, leading to different nonlinear
pictures of classical symmetries:
\begin{enumerate}
\item [i)] We assume that both functions $f$ and $g$ are analytic, and besides
\bel{11} \lim_{x\to\infty}f(x,y) \ = \ \lim_{x\to\infty} g(x,y) \  = \ \infty
\qquad \lim_{x\to\infty} \frac{f^2(x,y)}{g^2(x,y)} \ = \ a<\infty \ee In such a
case one obtains in the limit ${E\to\infty}$ \bel{12} \frac{\vec P^2}{\kappa^2
c^2} \ = \ \frac{\kappa^2 f^2 - m_0^2}{\kappa^2 g^2} \mathop{{\longrightarrow
}}\limits_{E \to \infty} a \ee or \bel{13} \lim_{E\to\infty}\vec{P}^2 \ = \
a\kappa^2 c^2 \ee Such type of theory is characterized at infinite energy by a
maximal value \r{13} of three-momenta. An example of such a theory is DSR1 of
Amelino-Camelia [1,2], which is obtained if for $a=1$ we choose the functions
$f$ and $g$ relating classical and bicrossproduct basis in $\kappa$-deformed
theory [19,20] \bel{14}f\left(E,\vec{P}{}^2\right) \ = \ \sinh\frac{E}{\kappa
c^2} + \frac{\vec{P}{}^2}{2\kappa^2 c^2} e^\frac{E}{\kappa c^2} \qquad
g\left(E,\vec{P}{}^2\right) \ = \ e^\frac{E}{\kappa c^2}\ee Of course there is
an infinite variety of theories having finite limit \r{13} for $E \to \infty$.
\item[ii)] Other type of theories are obtained if
we assume that $f$ and $g$ are singular for $E = \kappa c^2$, but in the limit
$E \to \kappa c^2$ the quotient $f/g$ is finite. Such a theory provides
divergent value of $E\0$ for $E=\kappa c^2$, because it maps the infinite range
of energies $E\0$ into finite interval $0<E<\kappa c^2$. A simple example of
such nonlinear transformation (see also [21]) \bel{15} f(E) \ = \
\frac{E}{\kappa c^2} \left( 1 - \frac{E}{\kappa c^2}\right)^{-1}\qquad g(E) \ =
\ \left( 1 - \frac{E}{\kappa c^2}\right)^{-1}  \ee provides the deformation of
Lorentz symmetry proposed recently by Magueijo and Smolin [3].

\end{enumerate}
In order to calculate the coproducts of deformed theories one should provide
the formulae inverse to the relations \r{8} (we use the notation $P\0\equiv
|\vec P\0|$): \bel{16}  E \ = \ \kappa c^2 F (E\0,P\0)\qquad \vec P \ = \ \vec
P\0 G (E\0, P\0) \ee Then we should use the formulae \bel{17}\bea{l} \cop(E) \
= \ \kappa c^2 F \left(\cop\0(E\0),\cop\0(P\0)\right)\\[3mm] \cop(\vec P) \ = \
\cop\0(\vec P\0)  G \left(\cop\0(E\0), \cop\0(P\0)\right)\ea \ee and reexpress
on rhs of \r{17} the operators $E\0$, $\vec P\0$ in terms of $E$, $\vec P$  by
means of the relations \r{8}.

As an example one can calculate the fourmomentum coproducts \r{17} for the
choice \r{14}, which is the energy-momentum dispersion relation for
bicrossproduct basis of $\kappa$-Poincar\'e algebra, further used as basic
postulate of DSR1 theory. The inverse formulae look in this case as follows:
\bel{18} F (E\0, P\0) \ = \ \ln D(E\0,  P\0) \qquad G (E\0,P\0) \ = \
D^{-1}(E\0, P\0)  \ee where \bel{19} D(E\0,P\0) \  = \ \frac{E\0}{\kappa c^2}
+\sqrt{ 1 + \frac{(E\0)^2}{\kappa^2 c^4}+ \frac{(\vec P\0)^2}{\kappa^2 c^2}}
\ee In linear approximation one gets \bel{20}E \ = \ E\0
-\frac{1}{2\kappa}(\vec P\0)^2 + O({1\over\kappa^2})\qquad \vec P \ = \ \vec
P\0\left(1-\frac{E\0}{\kappa c^2}\right) +O({1\over\kappa^2})\ee After the
approximation to the formula \r{8} with the choice \r{14} \bel{21} E\0 \ = \ E
+\frac{\vec{P}{}^2}{2\kappa}\left(1+\frac{E}{\kappa c^2}\right)+
O({1\over\kappa^2})\qquad \vec P\0 \  = \ \vec P\left(1+\frac{E}{\kappa
c^2}\right) +O({1\over\kappa^2}) \ee one gets the following deformation of the
primitive coproduct  \bel{22} \bea{l} \cop(E) \ = \ E\tens 1 + 1 \tens E -
\frac{1}{\kappa} \vec P \tens \vec P + O (\frac 1{\kappa^2})\\[3mm] \cop(\vec
P) \ = \ \vec P \tens 1 + 1 \tens \vec P - \frac{1}{\kappa c^2} (\vec P \tens E
+ E\tens \vec P) + O(\frac{1}{\kappa^2})\ea\ee
 Comparing the composition laws
\r{3},\r{4} for threemomenta and energy in DSR1 theory with \r{22} we see that
they follow from the nonlinear realization of classical Poin\-car\'e algebra.

 Further it can be shown that the $\kappa$-deformed boost transformations of
 DSR1 calculated in [17] can be obtained from classical Lorentz transformations
 after introducing the nonlinear change \r{16} of fourmomenta with the
 functions $F$ and $G$ given by \r{18}.

 We start with the formulae for finite classical Lorentz boosts
 (we choose $\vec{n}$ in the direction of relative velocities $\vec{v} = c
 \vec{n} \tanh\alpha$ of two Lorentz frames)\bel{23} \bea{l} E\0(\alpha) \ = \
 E\0 \cosh\alpha - c(\vec{n}\vec{P}\0) \sinh\alpha\\[3mm] \vec{P}\0 (\alpha) \ =
 \ \vec{P}\0 + \vec{n}\left[(\vec{n}\vec{P}\0) (\cosh\alpha - 1) -
\frac{E\0}{c}\sinh\alpha\right]\ea\ee and the relations \r{16}, \r{18} written
in the form \bel{24} D(E\0, P\0) \ = \ \exp(\frac{E}{\kappa c^2})\qquad \vec{P}
\ = \ D^{-1}(E\0, P\0) \vec{P}\0\ee
 where $D$ is given by \r{19}. Because\bel{25}\bea{l} E(\alpha) \ = \
 e^{i\alpha(\vec{n}\vec{N})} E e^{-i\alpha(\vec{n}\vec{N})} \ = \ \kappa c^2
 \ln D(E\0(\alpha), P\0(\alpha))\\[3mm] \vec{P}(\alpha) \ = \ e^{i\alpha(\vec{n}\vec{N})} \vec{P}
 e^{-i\alpha(\vec{n}\vec{N})} \ = \ D^{-1}\left(E\0(\alpha), P\0(\alpha)\right)
 \vec{P}\0 (\alpha)\ea\ee and\bel{26}\bea{l}D(E\0(\alpha),P\0(\alpha)) \ = \
  e^{i\alpha(\vec{n}\vec{N})} D(E\0, P\0) e^{-i\alpha(\vec{n}\vec{N})} =\\[3mm]
  = \ \frac{E\0(\alpha)}{\kappa c^2} + \sqrt{1+\frac{(E\0(\alpha))^2}{\kappa^2 c^4}
- \frac{(P\0(\alpha))^2}{\kappa^2 c^2}} =\\[3mm]= \ \frac{E\0(\alpha)}{\kappa
c^2} + \sqrt{1+\frac{(E\0)^2}{\kappa^2 c^4} - \frac{(P\0)^2}{\kappa^2
c^2}}=\\[3mm]= \ {1\over \kappa c^2}\left(E\0(\cosh\alpha - 1) -
c(\vec{n}\vec{P}\0)\sinh\alpha\right) + D(E\0, P\0) \ea\ee one can
write\bel{27}\bea{l} W(\alpha) \ = \ \exp\left(\frac{E(\alpha) - E}{\kappa
c^2}\right) \ =\\[3mm]= \ 1 + {1\over \kappa c^2}e^{-\frac{E}{\kappa
c^2}}\left[E\0(\cosh\alpha - 1) - c (\vec{n}\vec{P}\0)\sinh\alpha\right] \
=\\[3mm]= \ 1 + B(E)(\cosh\alpha - 1) - \frac{(\vec{n}\vec{P})}{\kappa c}
\sinh\alpha\ea\ee where\bel{28}B(E) \ = \ {1\over 2}\left(1 -
e^{-\frac{2E}{\kappa c^2}} + \frac{\vec{P}{}^2}{\kappa^2 c^2}\right)\ee From
\r{27} follows the boost transformations for energy
\begin{lit}\bel{29a}\exp\left(\frac{E(\alpha) - E}{\kappa c^2}\right) \ = \
W(\alpha) \Rightarrow E(\alpha) \ = \ E + \kappa c^2 \ln W(\alpha)\ee and for
the threemomentum $$\bea{l} \vec{P}(\alpha) \ = \vec{P}\0(\alpha)
e^{-\frac{E(\alpha)}{\kappa c^2}} \ = \ W^{-1}(\alpha) D^{-1}(E\0,
P\0)\vec{P}\0(\alpha) =\\[3mm]W^{-1}(\alpha) D^{-1}(E\0, P\0)\left(\vec{P}\0 +
\vec{n}\left[(\vec{n}\vec{P}\0)(\cosh\alpha - 1) - \frac{E\0}{c}
\sinh\alpha\right]\right)=\\[3mm]W^{-1}(\alpha)\left(\vec{P} +
\vec{n}\left[(\vec{n}\vec{P})(\cosh\alpha - 1) - \kappa c B(E)
\sinh\alpha\right]\right)\hfill(29b)\ea$$
\end{lit}
The formulae $(29a,b)$ in simpler version when $\vec{n}=(-1,0,0)$ were derived
by other method in [17] as the finite deformed boost transformations in DSR1
theory.

\section{$\kappa$-Deformed relativistic symmetries}

The quantum deformation of Poincar\'e algebra with dimensionfull deformation
parameter -- the fundamental mass parameter $\kappa$ -- has been introduced in
1991 in so-called standard basis [4-6]; further this deformation was rewritten
in bicrossproduct basis [7-11]. In bicrossproduct basis the Lorentz subalgebra
remains undeformed \bel{30} \bea{l} [M_i,M_j] \ = \ i \epsilon _{ijk} M_k
\\[3mm] [M_i,N_j] \ = \ i \epsilon _{ijk} N_k \\[3mm] [N_i,N_j] \ = \ -i \epsilon _{ijk}
M_k \\[3mm] \ea \ee where $\vec{M}=(M_1,M_2,M_3)$ generate space rotations, and
$\vec{N}=(N_1,N_2,N_3)$ the relativistic boosts. In the relations
  \r{5a}-\r{5c}
only the relation \r{5b} is deformed. One gets \bel{31}[N_i,P_j] \ = \  {i\over
2}\delta_{ij}\left[\kappa c\left(1 - e^{-\frac{E}{\kappa c^2}}\right)\right] -
\frac {i}{\kappa c} P_i P_j \ee
All remaining Poincar\'e algebra relations
remain classical.

The Casimir operator for the $\kappa$-deformed Poincar\'e algebra with deformed
relations given by (3.2) is given by formula \r{1}.

The Hopf algebra structure is provided by the following coproducts:
\begin{lit}
\begin{eqnarray}
\label{32a} \cop(E) & = & E\tens 1 + 1\tens E\\ \label{32b} \cop(\vec{P}) & = &
\vec{P}\tens 1 + e^{-\frac{E}{\kappa c^2}} \tens \vec{P}\\
\label{32c}\cop(\vec{M}) & = & \vec{M}\tens 1 + 1 \tens \vec{M}\\
\label{32d}\cop(N_i) & = & N_i\otimes 1 + e^{-\frac{E} {\kappa c^2}}\otimes N_i
+ {\frac{1}{\kappa c}}\epsilon_{ijk}P_j\otimes M_k\end{eqnarray}\end{lit} The
``quantum inverse'', the antipode, is given by the relations \bel{33} \bea{l}
S(E) \ = \ -E\\[3mm] S(\vec{P}) \ = \ -\vec{P} e^\frac{E}{\kappa c^2}
\\[3mm] S(\vec{M}) \ = \ -\vec{M}\\[3mm]
S(N_i) \ = \ -e^{\frac{E}{\kappa c^2}}N_i+{\frac{1}{\kappa
c}}\epsilon_{ijk}e^{\frac{E}{\kappa c^2}}P_j M_k\ea\ee We see that as Hopf
algebra only the subalgebra $(\vec{M},E)$ remains classical.

In order to obtain finite Lorentz transformations of fourmomenta which are
consistent with Hopf algebra structure one should introduce the finite boosts
in the pseudoEuclidean plane $(P_0, (\vec{n}\vec{P}))$ by the following formula
(see also [9], Sect.2d.):\bel{34} P_\mu (\alpha) \ = \
ad_{e^{i\alpha(\vec{n}\vec{N})}} P_\mu \ee with the quantum adjoint action
defined as follows (see e.g. [22]) \bel{35} ad_Y X \ = \ Y_{(1)} X
S(Y_{(2)})\ee and (in symbolic notation)
%\\
  $\cop(Y) = Y_{(1)}\otimes Y_{(2)}$.
Applying the formula $ad_{Y_1 Y_2}(X) = ad_{Y_1}(ad_{Y_2}(X))$ one
gets\bel{36}P_\mu(\alpha) \ = \ \sum_{k=0}^{\infty}
\frac{(i\alpha)^k}{k!}(\underbrace{ad_{(\vec{n}\vec{N})}(ad_{(\vec{n}\vec{N})}...
(ad_{(\vec{n}\vec{N})} P_\mu)...)}_{k-times})\ee Surprisingly enough,
substituting in \r{35} the coproduct and antipode from \r{32d}, \r{33}  for
arbitrary analytic function $h(P_0,P_1,P_2,P_3)$ one obtains \bel{37}
ad_{(\vec{n}\vec{N})} h(P_0, P_1,P_2,P_3) \ = \ [(\vec{n}\vec{N}), h(P_0,
P_1,P_2,P_3)]\ee i.e. one gets the "classical formula" describing finite boost
transformations for $\kappa$-deformed Poincar\'{e} algebra in bicrossproduct
basis:\bel{38} P_\mu(\alpha) \ = \ e^{i\alpha (\vec{n}\vec{N})} P_\mu\,
e^{-i\alpha (\vec{n}\vec{N})}\ee The differential equations following from the
relation \r{38} (for particular choice $\vec{n}=(-1,0,0)$) were used in [17]
for the calculation of finite boost transformations $(29a,b)$ in DSR1 theory.
We see therefore that the relation \r{38} can be obtained in two different
ways from the formula \r{36}
\begin{enumerate}
\item[i)] Using the Hopf algebra structure of $\kappa$-deformed Poincar\'{e}
algebra, as demonstrated above,
\item[ii)] By inserting in \r{36} the classical coproduct for the boost generators and
nonlinear formulae \r{16}, \r{18} describing threemomentum and energy.
\end{enumerate}

Using the following formula for coproducts [22]\bel{39} ad_{\cop X}\cop Y \ = \
(ad_{X_{(1)}} Y_{(1)})\otimes (ad_{\cop X_{(2)}} Y_{(2)})\ee one can also
derive \footnote{In particular, $ad_{N_k} P_i=[N_k,P_i]\Rightarrow
\Delta(ad_{N_k} P_i)=[\Delta N_k,\Delta P_i]\neq [N_k\otimes 1+1\otimes
N_k,\Delta P_i]=ad_{\Delta N_k}\Delta P_i$. The equality sign holds for the
primitive coproduct $\Delta\vec{N}=\vec{N}\otimes 1+1\otimes \vec{N}$, i.e. for
nondeformed boosts. In other words, $\kappa$-deformed momentum coalgebra is not
the left (or right) Lorentz-module coalgebra (see [22,23])}\bel{40} ad_{\cop
N_k}\cop P_\mu \ = \ [N_k\otimes 1 + 1\otimes N_k, \cop P_\mu]\ee where we
recall that for $\kappa$-deformed Poincar\'{e} algebra the coproduct for the
boosts $N_i$ and threemomenta $P_i$ is nonprimitive (see $(32b,d)$). Using
\r{40} one can show that in general
\begin{lit}
\bel{41a}\bea{l}\vec{P}_\Delta(\alpha) \ \equiv \
ad_{e^{i\alpha(\vec{n}\cop\vec{N})}} \cop \vec{P} \ =\\[3mm]= \
e^{i\alpha(\vec{n}\vec{N})}\otimes
e^{i\alpha(\vec{n}\vec{N})}\left(\vec{P}\otimes 1 + e^{-\frac{E}{\kappa
c^2}}\otimes \vec{P}\right) e^{-i\alpha(\vec{n}\vec{N})}\otimes
e^{-i\alpha(\vec{n}\vec{N})}=\\[3mm] = \ \vec{P}(\alpha)\otimes 1 +
e^{-\frac{E(\alpha)}{\kappa c^2}}\otimes
\vec{P}(\alpha)\ea\ee\bel{41b}\bea{l}E_\Delta(\alpha) \ \equiv \
ad_{e^{i\alpha(\vec{n}\cop\vec{N})}} \cop \vec{E} \ =\\[3mm]= \
e^{i\alpha(\vec{n}\vec{N})}\otimes e^{i\alpha(\vec{n}\vec{N})}\left(E\otimes 1
+ 1\otimes E \right) e^{-i\alpha(\vec{n}\vec{N})}\otimes
e^{-i\alpha(\vec{n}\vec{N})} =\\[3mm]= \ E(\alpha)\otimes 1 + 1\otimes
E(\alpha)\ea\ee
\end{lit}

\noindent i.e. the quantum boost
 transformations $(29a,b)$ for the coproduct of
four-momenta are consistent with the $\kappa$-deformed addition law $(32a,b)$.

One can conclude that for the deformed relativistic energy-momentum dispersion
relation \r{1} there are two consistent addition laws for deformed
four-momenta, one symmetric (cocommutative) and second noncocommutative.
\begin{enumerate}
\item[i)] The symmetric addition law is the result of nonlinear transformations
of the three-momentum and energy generators in classical framework of
relativistic symmetries. One obtains nonabelian symmetric addition law for both
energy and three-momenta. Due to this symmetry (cocommutativity of the
coproduct) the dual Minkowski space will be a standard one, with standard
commutative space-time coordinates.

\item[ii)] The  nonsymmetric  ( noncocommutative ) addition law for deformed
three\-mo\-menta permits in the framework of $\kappa$-deformed quantum group to
obtain the abelian addition law for energy. Due to noncocommutativity of the
coproduct in the fourmomentum space, one obtains from Hopf algebra duality
($<x_\mu,P_\nu>=ig_{\mu\nu}$) [22] the $\kappa$-deformed space-time [7-11]
($i=1,2,3$)\bel{42} [x_0, x_i] \ = \ {i\over \kappa c}x_i\qquad [x_i, x_j] \ =
\ 0\ee\end{enumerate}If we are not requiring the classical additivity of energy
one can consider any other bases for fourmomenta preserving bicrossproduct
property. In particular

- one can describe $\kappa$-deformed framework with classical Poincar\'{e}
basis [19] - the nonabelian addition law for energy [20] can be interpreted as
introducing geometric 2-particle interactions in standard relativistic theory.

- by performing the nonlinear transformations of fourmomenta it is possible to
describe the $\kappa$-deformed Poincar\'{e} algebra in the basis which provides
deformed energy-momentum dispersion relation given by Magueijo and Smolin [3],
singular for energy $E=\kappa c^2$ (see [21]).

\section{Conclusions}

It follows from our discussion that the introduction of large class of deformed
energy-momentum dispersion relations (see \r{10}) preserved under deformed
Lorentz transformations is easy to achieve by a nonlinear change of classical
fourmomentum operators (see \r{8} and \r{16}). In such a framework
 all these nonlinearly related
symmetry schemes are described by the same Hopf algebra -- a mathematical tool
for description of given algebraic symmetry. The real distinction between
different deformed relativistic symmetry schemes is due to the choice of
coproduct, in particular describing the composition law of energy-momentum for
multiple states. The nontrivial deformations are obtained if the coproduct is
not symmetric. The classification of nontrivial quantum deformations of
Poincar\'{e} symmetries is known [24] and it is almost complete. From physical
reasons there are distinguished  these deformations which preserve the commutativity of
fourmomentum generators. In such a case it follows from the duality of Hopf
algebras describing space-time coordinates and fourmomenta that there are
possible only Lie - algebraic deformations of Minkowski space-time. Assuming
that we keep classical $O(3)$ invariance only the following two deformations
introducing fundamental mass parameter are possible:\begin{enumerate}\item[i)]
$\kappa$-deformed Minkowski space, described by the relations \r{42}. Such
deformation is physically interesting, because it preserves undeformed
nonrelativistic symmetries and the effective deformation parameters $P_0/\kappa
c$ and $\vec{P}^2/\kappa^2 c^2$ shows that $\kappa$-deformation is a high
energy / high momentum effect.

\item[ii)] other possible deformation covariant under classical $O(3)$
rotations is provided by relations:\bel{43}[\hat{x}_0,\hat{x}_i] \ = \ 0\qquad
[\hat{x}_i,\hat{x}_j] \ = \ {i\over \kappa} \varepsilon_{ijk} \hat{x}_k\ee The
second formula \r{43}, used as a description of fuzzy sphere (see e.g. [25,26])
modifies the nonrelativistic sector of classical relativistic symmetries, i.e.
one can argue that this second choice for physical applications is less
attractive.\end{enumerate}

We would like to stress that we are aware of difficulties with the adjustment
to physical interpretation of nonsymmetric coproducts. It is also true that the
complete structure of quantum group sometimes requires more sophisticated
mechanisms to detect the deformation effects (see e.g. [27]). We would like
however to point out that whatever are the efforts to accommodate the
nonsymmetric coproducts \r{32b} into the description of physical processes (see
e.g. [28]), at present the alternative for DSR theories is clear: or classical
Einsteinian framework or quantum $\kappa$-deformed symmetry, and declaration
should be made. Also in
 discussions of modified relativity principles
 two formal steps which are
sometimes mixed -- the nonlinear change of basis and the notion of deformation
-- should be clearly separated.

\subsection*{Acknowledgement}
One of the authors (J.L.) would like to thank Piotr Kosi\'{n}ski for valuable
discussion and Jerzy Kowalski-Glikman for information about
  the  ref. [21].
%%%%\newpage

\end{document}